\documentclass{article} %
\usepackage{iclr2021_conference,times}

\usepackage{amsmath,amsfonts,bm}

\def\eqref#1{equation~\ref{#1}}

\def\1{\bm{1}}

\DeclareMathAlphabet{\mathsfit}{\encodingdefault}{\sfdefault}{m}{sl}
\SetMathAlphabet{\mathsfit}{bold}{\encodingdefault}{\sfdefault}{bx}{n}

\usepackage{graphicx}

\usepackage[super]{nth}
\usepackage{subcaption}
\usepackage{booktabs}
\usepackage{verbatim}
\usepackage[switch]{lineno}
\usepackage{url}
\usepackage{hyperref}
\usepackage{pifont}%
\hypersetup{
  colorlinks=true,
  urlcolor=blue,
  citecolor=black
}
\newcommand*\samethanks[1][\value{footnote}]{\footnotemark[#1]}

\title{Impact of data-splits on generalization:\\Identifying COVID-19 from cough and context}

\author{Makkunda Sharma\thanks{Equal contribution}, Nikhil Shenoy\samethanks[1] , Jigar Doshi\samethanks[1] , Piyush Bagad, Aman Dalmia \\
Wadhwani Institute for Artificial Intelligence\\
\texttt{\{makkunda,shenoy,jigar,piyush\}@wadhwaniai.org} \\
\And
Parag Bhamare, Amrita Mahale, Saurabh Rane, Neeraj Agrawal \& Rahul Panicker \\
Wadhwani Institute for Artificial Intelligence \\
}

\iclrfinalcopy %
\begin{document}

\maketitle

\begin{abstract}
Rapidly scaling screening, testing and quarantine has shown to be an effective strategy to combat the COVID-19 pandemic. We consider the application of deep learning techniques to distinguish individuals with COVID from non-COVID by using data acquireable from a phone. Using cough and context (symptoms and meta-data) represent such a promising approach. Several independent works in this direction have shown promising results. However, none of them report performance across clinically relevant data-splits. Specifically, the performance where the development and test sets are split in time (retrospective validation) and across sites (broad validation).  Although there is meaningful generalization across these splits the performance significantly varies (up to 0.1 AUC score) [Table 1]. In addition, we study the performance on symptomatic and asymptomatic individuals across these three splits. Finally, we show that our model focuses on meaningful features of the input, ‘cough’ bouts for cough and relevant symptoms for context. The code and checkpoints are available at : \href{https://github.com/WadhwaniAI/cough-against-covid}{https://github.com/WadhwaniAI/cough-against-covid} \footnote[1]{The data will be available at this link when all relevant permissions are granted.}

\end{abstract}

\section{Introduction}
\label{sec:intro}

 Screening, testing and quarantining has been an effective strategy employed by public health systems around the world to combat the spread of the COVID-19. There are two main objectives of any public health system, namely, minimize the spread of the virus and save lives. To achieve these objectives, especially during a pandemic, there are limited resources available like testing kits, hospital beds, x-ray machines etc. We consider the approach of using machine learning algorithms to increase the efficiency of resource allocation within the public health system. Furthermore, a non-invasive, widely accessible, scalable, and accurate model could be an impactful solution. Identifying COVID-19 from respiratory acoustics like cough and user-reported symptoms represent two such approaches. Both of these approaches simply require a phone with a microphone. This could be used by the public health system as a screening tool for effective allocation of the lab-tests. 

There has been several prior work using cough (\cite{brown2020exploring, Imran2020-rf, Bagad2020-rx}), voice (\cite{Bartl-Pokorny2020-ob, Pinkas2020-pd}), breathing sound (\cite{Faezipour2020-rw}), symptoms and their combinations (\cite{Schuller2020-gj, Coppock2021-ez}) to identify COVID-19. \cite{Qian2020-oa} and \cite{Deshpande2020-on} provide fairly comprehensive surveys on related works. Given this evidence and the approaches’ practical advantages, it’s surprising that it has seen limited clinical deployment. One of the reasons could be the lack of trust from the healthcare community as verbalized by \cite{Topol2020} in this article. 
This paper debates highly publicized work claiming 98.5\% accuracy for identifying COVID from cough. Their critical review could be summarized as following. 1) Lack of reproducibility 2) Lack of clinically relevant evaluations 3) Interpretation of deep learning models. In this work, we address these limitations to build confidence in this line of work and spur further research. 

The main contribution of this work is as following:

1) Dataset and model: We describe our multi-site, lab-tested ground-truth dataset collected for this paper. Detailed description of our modelling approach to identify COVID-19 from cough and contextual-data.

2) Evaluations: We show the model’s performance across three clinically meaningful data splits as defined in this paper by \cite{Norgeot2020} and \cite{Kleppe2021}. Specifically, the performance where the development and test sets are split randomly, in time (retrospective validation) and across sites (broad validation). We see meaningful generalization across these splits however the performance varies up to 0.1 AUC score [Table \ref{tbl:res}]. 
In addition, we report model performance across a special sub-populations: symptomatic and asymptomatic individuals [Table \ref{tbl:res}].

3) Interpretability: We compute the saliency maps of the cough classifier and show qualitatively it focuses on the ‘cough’ parts of the input. For context-based classifiers, we show the most predictable features at an instance level.

\begin{figure}
\centering
\begin{subfigure}{.6\textwidth}
  \centering
  \includegraphics[width=\linewidth]{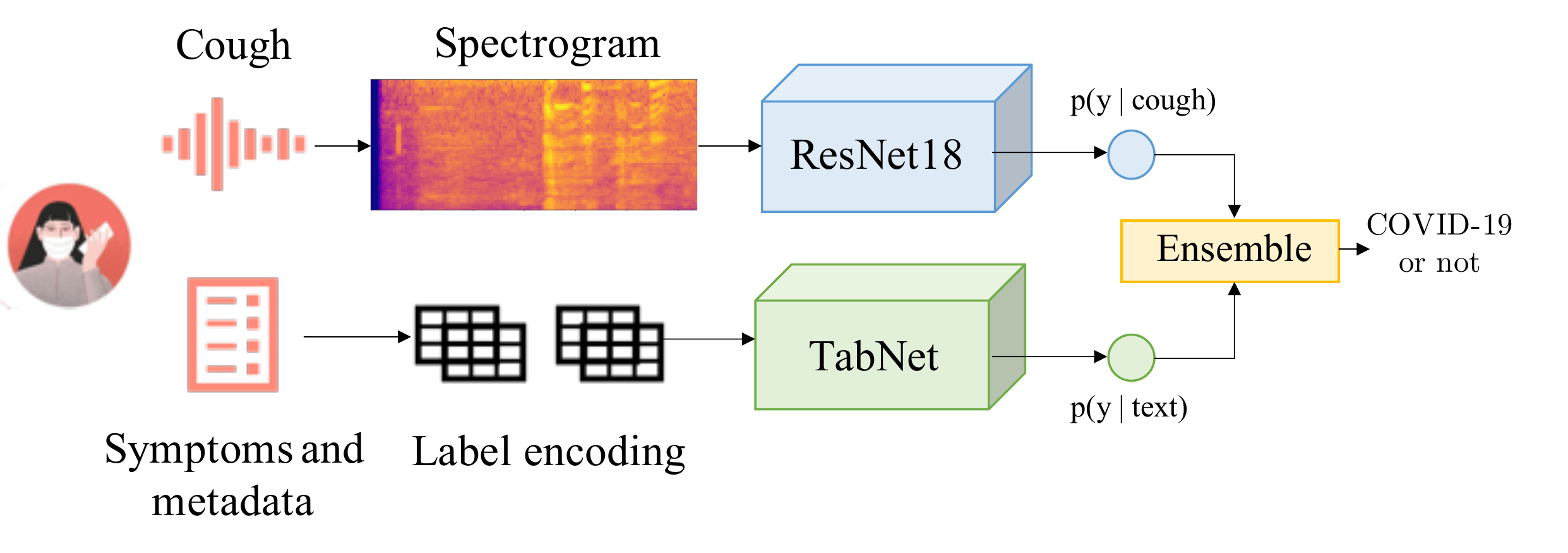}
  \caption{\textit{Schematic diagram of the proposed solution }}
  \label{fig:arch}
\end{subfigure}%
\begin{subfigure}{.4\textwidth}
  \centering
  \includegraphics[width=\linewidth]{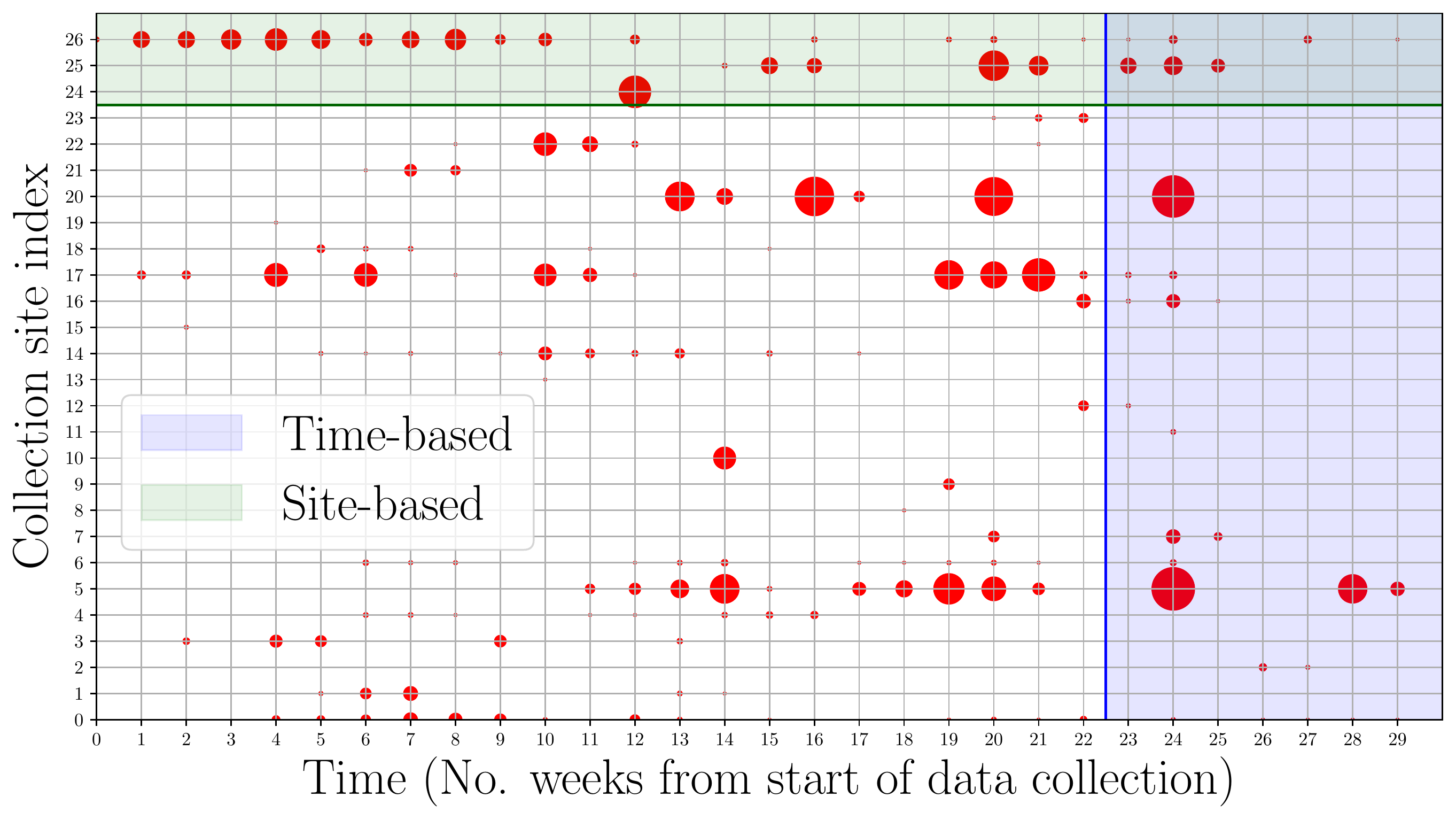}
  \caption{\textit{Data splitting strategy}}
  \label{fig:data_slice}
\end{subfigure}
\caption{(a) For every individual, we get predicted probabilities $p(y = 1| \mathbf{x}_{\text{cough}})$ and $p(y = 1| \mathbf{x}_{\text{context}})$ from the individual classifiers. For our final prediction we use a simple ensembling scheme that averages the predictions from the two classifiers. (b)  The data has been partitioned into train/test splits in 3 ways: (i) random, (ii) time-based (blue region denotes the test set), (iii) site-based (green).  The size of a point denotes the number of samples collected in a given week ($x$-axis) at a given site ($y$-axis) of the data collection drive.}
\label{fig:test}
\end{figure}

\section{Data}

\noindent\textbf{Open-source non-COVID cough datasets:}
\label{sec:pretrained-datasets}
Given the challenges of training deep learning model from scratch on a relatively small dataset, we collate a larger dataset of audio samples from various public datasets - FreeSound Database by \cite{fonseca2018general}, FluSense by \cite{al2020flusense} and Coswara by \cite{sharma2020coswara} which we use to pretrain our model for cough detection. In total we obtain 31,909 sounds segments, of which 27,116 are non-cough respiratory sounds (wheezes, crackles or breathing) or human speech, and 4,793 are cough sounds.

\noindent\textbf{COVID-19 dataset:}
We collect this dataset from individuals who have undergone a COVID-19 test, from numerous testing sites across the country. In addition, contextual data such as symptoms, travel history, contact with confirmed case and demographic information etc is collected. Unlike crowd-sourced datasets such as \cite{brown2020exploring}, \cite{coughvid} and \cite{sharma2020coswara} that rely on self-reported COVID-19 status, our ground-truth is lab test results from the healthcare facilities.
This dataset consists of 12,780 cough sounds from 4,260 individuals from 27 different sites. 1,394 have a positive test result and 2866 remaining are tested negatives. Please see Appendix \ref{appendix-data-collection}, \ref{appendix-contextual} for more details on data collection process and list of contextual features.

\noindent\textbf{Dataset splits:} A key challenge in developing deep learning models on real-world data is to devise the right developmental (train-validation) and test splits. In order to establish robustness towards deployment of our model, we use 3 different strategies for splitting the data. All samples of an individual are always part of the same set. The details on the data distribution across the splits are shown in Figure \ref{fig:data-dist}.
\label{sec:data-split}

\noindent\textbf{1. Random}
We randomly split the data with train-validation-test splits being 80\%:10\%:10\%.

\noindent\textbf{2. Time-based}:
In order to perform a \textit{retrospective validation} as laid out by \cite{Kleppe2021}, we split the data by time - we select two cut-off dates s.t. the validation and test sets are about 10\% each and the remaining data forms the training set. This is demonstrated in Figure \ref{fig:data_slice} with $x$-axis denoting the time in weeks from the start of data collection.

\noindent\textbf{3. Site-based}:
In order to perform a \textit{broad validation} as laid out by \cite{Kleppe2021}, we split the data by sites. We construct the test set by selecting a set of 3 sites s.t. they constitute about 20\% of the dataset. This is demonstrated in Figure \ref{fig:data_slice} with $y$-axis denoting site indices in data collection.

\begin{table}[]
\centering
\begin{tabular}{lccc|ccc}
\toprule
Model         & \multicolumn{1}{l}{Task 1} & \multicolumn{1}{l}{Task 2} & \multicolumn{1}{l|}{Task 3} & \multicolumn{1}{l}{Task 1 - S/A} & \multicolumn{1}{l}{Task 2 - S/A} & \multicolumn{1}{l}{Task 3 - S/A} \\
\midrule
Cough-based   & 0.787                      & 0.690                      & 0.761                       & 0.820 / 0.713                                    &  0.510 / 0.699   &      0.709 / 0.613                                  \\
Context-based & 0.718                      & 0.650                      &   0.669                     & 0.610 / 0.730       &                 0.449 / 0.560             &        0.559 / 0.645                             \\
Ensembling    & \textbf{0.797}                      & \textbf{0.718}                      & \textbf{0.774}                       & 0.816 / 0.740                               &      0.498 / 0.709               &     0.707 / 0.671           \\
\bottomrule
\end{tabular}
\caption{Given the skewed label distribution, we use a distribution-agnostic metric: Receiver Operating Characteristic - Area Under Curve (AUC) (Left) Results on the test set for each of the 3 splits: Task 1 (random split), Task 2 (time-based split) , and Task 3 (site-based split). 
(Right) We analyze performance across symptomatics (S)/ asymptomatics (A) on all tasks. We consider an individual symptomatic if they report at least one of the symptoms: cough, fever, or shortness of breath.}
\label{tbl:res}
\end{table}

\section{Method}
\label{sec:method}

\subsection{Cough-based Classification}
Inspired by the recent success of CNNs applied to audio inputs by \cite{hershey2016cnn}, we develop a CNN-based framework that ingests spectrogram representations of audio and directly predicts the probability of the presence of COVID-19.
In this sections, we outline details of the input processing, model architecture, training strategies employed and inference for cough-based model. 

\noindent\textbf{Input processing:}
During training, we randomly sample a 2-second audio-segment of the cough recording and transform it  into a log-melspectrogram patch of $64 \times 201$ bins that forms the input to the classifier. We denote this input as $\mathbf{x}_{\text{cough}}$ and the predicted label as $y$. Details in Appendix \ref{sec:input-preprocessing}

\noindent\textbf{Data augmentations:}
In order to increase robustness to noise, we apply the following augmentations (online) while training: (a) addition of external background environmental sounds from ESC-50 dataset by \cite{piczak2015esc}, and (b) time and frequency masking of the spectrogram input as in  \cite{park2019specaugment}. At train time, we randomly select a noise sample from ESC-50, modulate the amplitude by a random factor between 0.4 and 0.75 and add it to the input cough sound. 

\noindent\textbf{Training:}
We use ResNet-18 by \cite{he2016deep} as the backbone network, followed by two linear layers and the final output layer with softmax activation that outputs the distribution $p(y| \mathbf{x}_{\text{cough}})$. Dropout with probability $0.4$ and ReLU activation are used after all linear layers.  

Our model is first pretrained on the open source cough datasets outlined in Sec. \ref{sec:pretrained-datasets}. We train the model to classify cough (not covid cough). 
For the downstream cough classification, we use this pretrained model to train with AdamW, an initial learning rate of $1e-4$ and a decay of 0.95 after every 10 epochs. We use a batch size of 128 and train for a total of 200 epochs. For evaluation, we pick a checkpoint based on the epoch with best AUC on validation set. We set the same seed for all experiments. The same training procedure is followed for all data-splits.

\noindent\textbf{Inference}:
Every cough sample is divided into 2-second segments using a sliding window with a hop length of 500ms. We pad inputs less than 2 seconds with zeros. We take the median
over the \textit{softmax} outputs for all the segments to obtain the prediction for a single sample. Since we have three cough samples per individual, we aggregate the predicted probabilities at an individual level using the \textit{max} operator. All performance metrics henceforth have been reported at the individual level.

\subsection{Context-based Classification}
Tree-based models have traditionally worked well when compared to deep neural networks (DNNs) on tabular datasets.
However, TabNet, as proposed by \cite{arik2019TabNet}, combines the benefits of gradient-based learning of DNNs and sparse feature selection of tree-based models. For our context-based classification task, we leverage TabNet as our classifier.

\noindent\textbf{Input Preprocessing:}
The input to the context-based classifier is a feature vector consisting of continuous variables (age, temperature etc) and categorical variables (presence of cough as symptom, travel history etc). For a detailed list of features used, refer to Appendix \ref{sec:contextual-features}. Categorical variables are label encoded while continuous variables are normalized to zero mean and unit standard deviation. We denote this input to the context-based classifier as $\mathbf{x}_{\text{context}}$.

\noindent\textbf{Modelling:} 
TabNet, a deep network for tabular data that uses sequential attention to choose which features to reason from at each decision step.
Input is a preprocessed context-based features and outputs two logits with softmax activation that produce the distribution $p(y| \mathbf{x}_{\text{context}})$. We train with the standard cross-entropy loss and use SGD optimizer with a learning rate of $1e-2$ and batch size of 128.
The validation set is used for early stopping, with AUC as the performance measure. All hyperparameters are set to default values used in this implementation. \footnote{\href{https://github.com/dreamquark-ai/TabNet}{https://github.com/dreamquark-ai/TabNet}}

\begin{figure}
\centering
\begin{subfigure}{.33\textwidth}
  \centering
    \includegraphics[width=\linewidth]{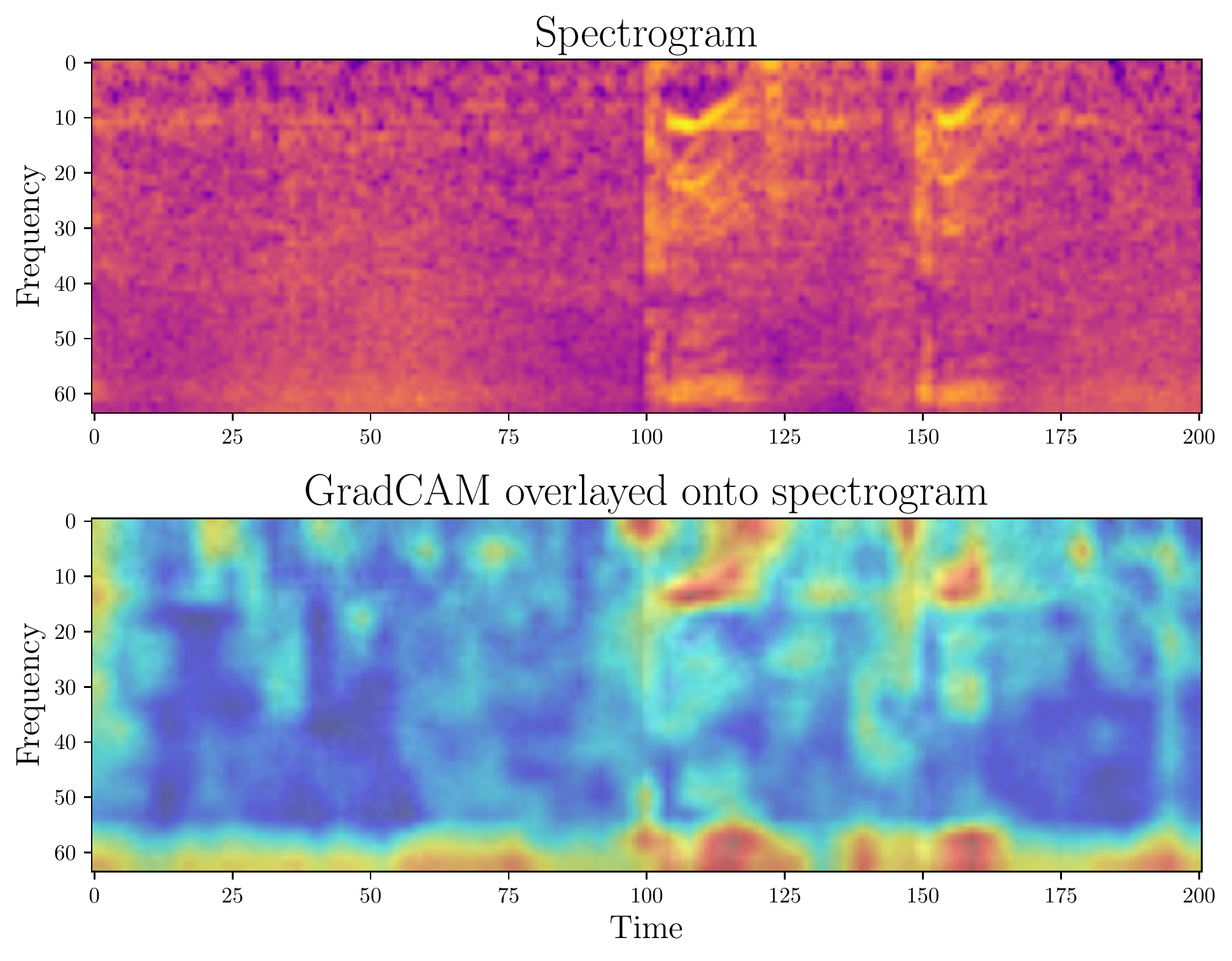}
  \caption{GradCAM++ saliency mask}
  \label{fig:gradcam}
\end{subfigure}%
\begin{subfigure}{.67\textwidth}
  \centering
    \includegraphics[width=\linewidth]{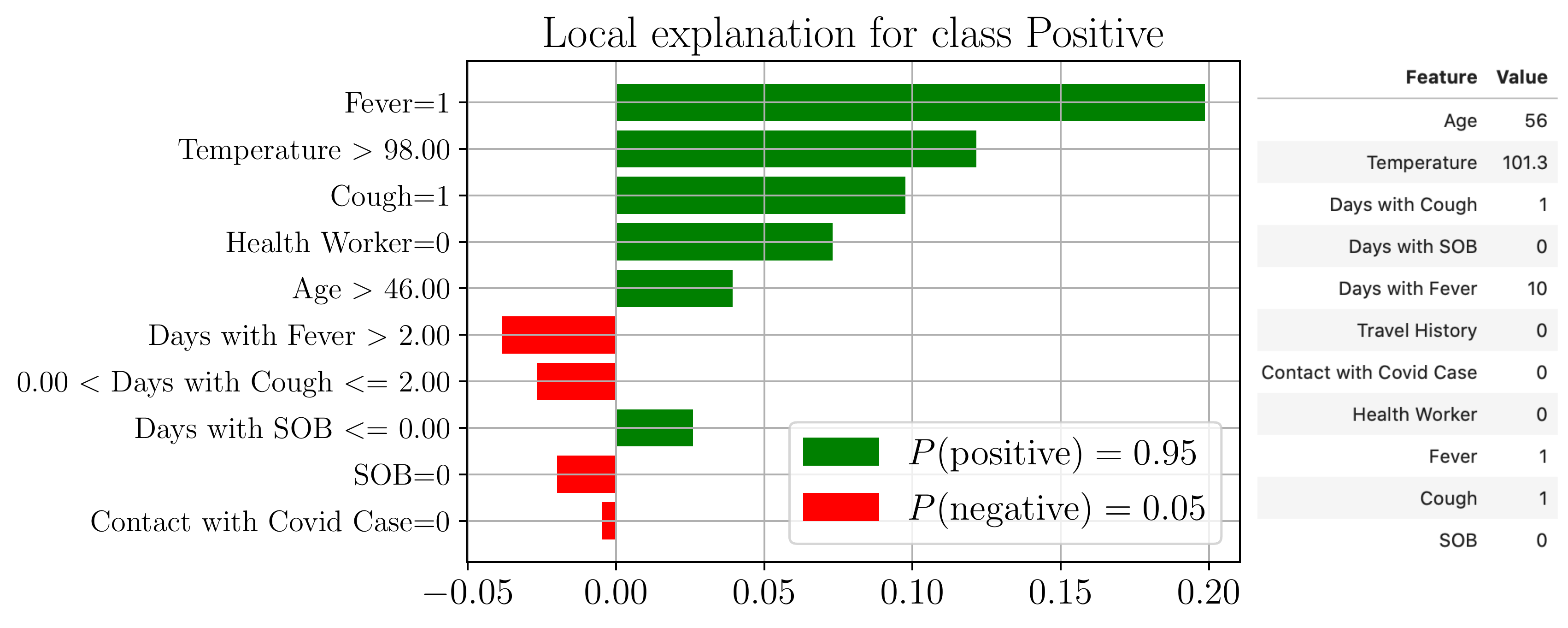}
  \caption{Contributions of context-based features to predictions}
  \label{fig:lime}
\end{subfigure}
\caption{(a) Interpretability for cough-based model: GradCAM++ saliency map of areas of focus of the model on input spectrogram image. 
(b) Interpretablity for context-based model: We analyze contributions of context-based features to the prediction score using the LIME method.
}
\label{fig:interpret}
\end{figure}

\section{Experimental Evaluation and Discussion}
\label{sec:Discussion}
\vspace{-1em}

\noindent\textbf{Performance across data splits}:
We train three models using the same hyper parameters as defined in \hyperref[sec:method]{Methods} for each of the three tasks/split. As evident from Table \ref{tbl:res}, we observe that it is easiest to generalize on the \textit{random} split followed by \textit{time-based} and \textit{site-based}. We hypothesize that this difference arises from the shift in label distribution across the splits. This also highlights the importance of selecting splits carefully since high-performant models on a randomized split may not generalize well in a deployment setting. Additionally, note that even though the performance on the other two splits is lower, they closely mimic the deployment scenario and thus it may induce confidence and are in-line with recommendations of \cite{Norgeot2020} and \cite{Kleppe2021}.

\noindent\textbf{Performance across (a)symptomatics}:
On the \textit{random} split, we report model performance across special sub-populations: symptomatic and asymptomatic groups in Table \ref{tbl:res}.
The cough-based model performs well for the two sub-groups, importantly, even for completely asymptomatic individuals.
Contrary to intuition, the context-based model performs much worse on symptomatics than on asymptomatics.
On further investigation, we find that the context-based model naively assigns lower probability scores for asymptomatic patients. Note that our dataset is skewed towards asymptomatics and the label distribution among them favors COVID-19 negative set. We suspect that the context-based model picks this association of asymptomatics being negative which disproportionately hurts its performance on symptomatics. The similar performance of cough-based model on these sub-populations is desirable since it is less prone to this data distribution skew.

\noindent\textbf{Interpretability}:
Given the clinical uncertainty of this task and the use of deep learning for it, it is essential for clinicians to qualitatively understand the model behaviour and its predictions. As a sanity check, we employ GradCAM++ by \cite{DBLP:journals/corr/abs-1710-11063} to compute these saliency map. We consistently observe that the focus-areas are on and around the cough bouts as shown in Figure \ref{fig:gradcam}. This reinforces the belief that the model is indeed making predictions based on cough signal and otherwise.

For the context-based classifier, we use Local Interpretable Model-agnostic Explanations (LIME by \cite{ribeiro2016should}) to understand which specific features help the model differentiate between COVID+ and COVID- patients at an instance level. For example in Figure \ref{fig:lime}, note that presence of fever/cough, temperature $> 98$ contribute to positive score whereas no shortness-of-breath contributes to a negative score. For more examples refer to \ref{app-interpret}.

\section*{Acknowledgements}
We greatly appreciate the support of our team members Nikhil Velpanur, Akshita Bhanjdeo, Patanjali Pahwa, Puskar Pandey, Bhavin Vadera, Vishal Agarwal, and Kalyani Shastry at Wadhwani AI. We would also like to thank our research partners James Zhou at Stanford, Peter Small and Puneet Dewan at Global Health Labs, and Ankit Baghel, Anoop Manjunath and Arda Sahiner from Stanford. We would like to acknowledge the generous support of our donors United  States Agency for International Development, Bill and Melinda Gates Foundation (BMGF), and AWS for GPU credits.

We would like to thank our data collection partners, the Governments of Bihar and Odisha, Municipal Corporation of Greater Mumbai, Ashfaq Bhat and his team at Norway India Partnership Initiative, Ravikant Singh and his team at Doctors for You, Pankaj Bhardwaj and Suman Saurabh from Department of Community Medicine in All India Institute of Medical sciences, Jodhpur and we would also like to thank our numerous field staff for their passion, hard work, and dedication. They have ensured strict adherence of the safety protocols through the data collection effort while maintaining high data quality during an active global pandemic.

\bibliography{iclr2021_conference}
\bibliographystyle{iclr2021_conference}

\appendix
\section{Appendix}

\subsection{Data collection details}
\label{appendix-data-collection}
Our data collection pipeline consists of the following stages: (i) collection of individual specific metadata, (ii) recording of audio samples and finally (iii) obtaining the results of the COVID-19 RT-PCR test. We manually clean and verify all collected data. We achieve this through three separate application interfaces. The details of data collected through these apps are enlisted below:

\begin{itemize}
    \item \textbf{Personal and Demographic information}: We collect the individual's name, mobile number, age, location (facility) and self-reported biological sex. 
    \item \textbf{Health-related information}: We collect the COVID-19 RT-PCR test result, body temperature and respiratory rate. We also note the presence of symptoms like fever, cough, shortness of breath and number of days since these symptoms first appear, and any measures undertaken specifically for cough relief. Finally, we also ask individuals if they have any co-morbidities. 
    \item \textbf{Additional metadata}: Additional data collected includes location (name of the facility, City and State), travel history of the individual, information about contact with confirmed COVID-19 cases, whether they are a health worker, and information about habits such as smoking, tobacco. 
\end{itemize}

\begin{figure}[!htbp]
    \centering
    \includegraphics[width=\linewidth]{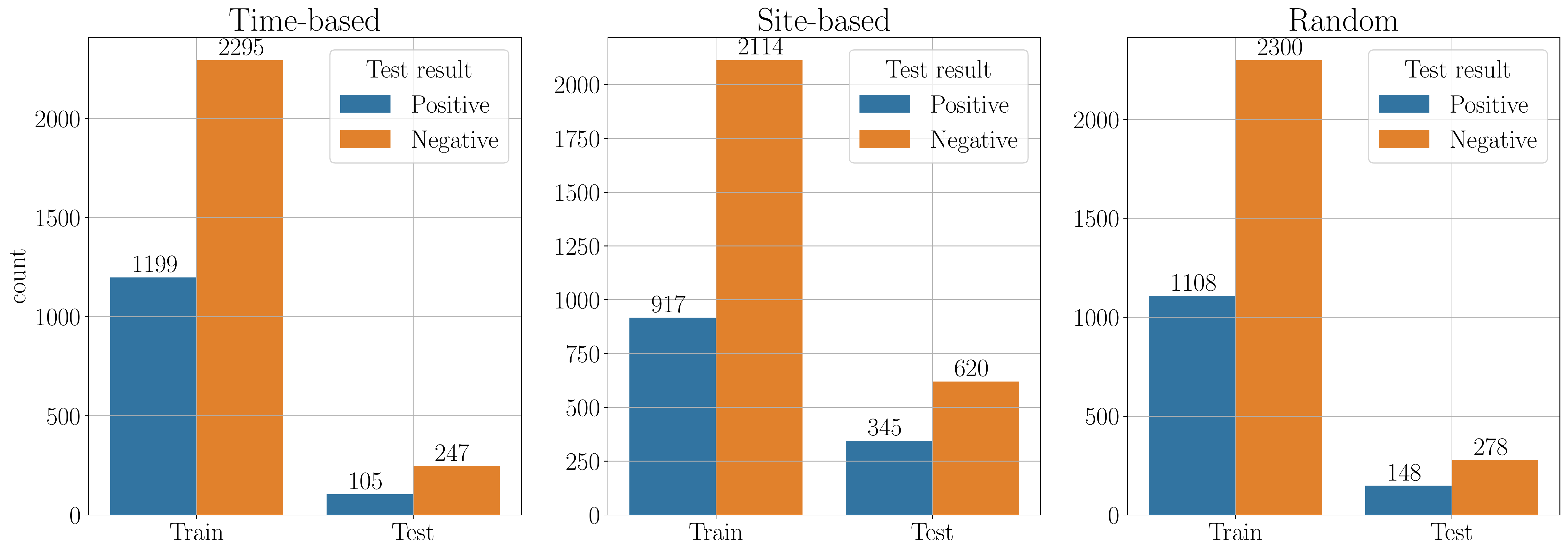}
    \caption{Distribution of positives and negatives in the train and test sets for each of the data splits.}
    \label{fig:data-dist}
\end{figure}

The distribution of positives and negatives in each of the splits are shown in Figure \ref{fig:data-dist}. The $x$-axis denotes the set (train or test) and $y$-axis denotes the number of individuals. Note that each individual has three cough samples thus the overall dataset size for cough-based classification triples.

\subsection{Contextual Features}
\label{sec:contextual-features}
The list of all features used in the context-based classification is provided in Table \ref{tab:context-features}
\label{appendix-contextual}
\begin{table}[!h]
\small
\centering
    \begin{tabular}{lll}
    \toprule
        Feature &  Type &  Other Details   \\ 
       \midrule
        Patient Age &  Continuous & - \\ 
        Patient Temperature &  Continuous & - \\ 
        Days with Cough &  Continuous & - \\ 
        Days with Shortness of Breath &  Continuous & - \\ 
        Days with Fever &  Continuous & - \\
        Presence of Cough &  Discrete & Yes/No \\
        Presence of Shortness of Breath &  Discrete & Yes/No \\
        Presence of Fever &  Discrete & Yes/No \\
        Contact with Covid Confirmed Case &  Discrete & Yes/No \\
        Is Patient a Health Worker &  Discrete & Yes/No \\
        Travel History &  Discrete &  No / Inter-district / Inter-state / Inter-country\\ \bottomrule

\end{tabular}
  \caption{Input features to context-based classifier}
  \label{tab:context-features}
\end{table}

\subsection{Details of Input Pre-processing}
\label{sec:input-preprocessing}
During training, we randomly sample a 2-second audio-segment of the input sample
and use short-term magnitude spectrograms as input to our CNN model. All audio is first converted to single-channel, 16-bit streams at a 16kHz sampling rate for consistency. Spectrograms are then generated in a sliding window fashion using a hamming window of width 32ms and hop 10ms with a 512-point FFT. This gives spectrograms of size $257 \times 201$ for 2 seconds of audio.
and is further integrated into 64 mel-spaced frequency bins with minimum frequency 125Hz and maximum frequency 7.5KHz, and the magnitude of each bin is log transformed.
This gives log-melspectrogram patches of $64 \times 201$ bins that form the input to all classifiers. Finally, the input is rescaled by the largest magnitude over the training set to bring the inputs between -1 and 1.

\subsection{Computation of Label Noise Probabilities}
\label{appendix-label}
Let the actual COVID-19 status be denoted by a random variable $C$ and that from RT-PCR be denoted by $R$. Let $S_n$ denote sensitivity and $S_p$ specificity. Also, let's assume 10\% prevalence i.e. $P(C = 0) = 0.9$.  We have
\begin{equation}
    S_n = \frac{P(R=1, C=1)}{P(C = 1)} = P(R=1| C=1) = 0.70;
\end{equation}
\begin{equation}
    S_p = \frac{P(R=0, C=0)}{P(C = 0)} =  = P(R=0| C=0) = 0.95
\end{equation}
Thus, we get $P(R = 0 | C = 1)=1-0.7= 0.3$ and $P(R = 1 | C = 0) = 1 - 0.95 = 0.05$ and $P(R = 0) = P(R = 0| C = 0) P(C = 0) + P(R = 0| C = 1) P(C = 1) = 0.885$ Applying the Bayes' rule, we get
\[
P(C = 0| R = 0) = \frac{P(R = 0| C = 0) P(C = 0)}{P(R = 0)} = 0.966
\]
Likewise, $P(C = 1| R = 0) = 0.033, P(C = 0| R = 1) = 0.3913, P(C = 1| R = 1) = 0.6087$. In reference to notation introduced earlier, we get 

\begin{equation}
    p_{\text{flip}}(l = 0) = P(C = 1 | R = 0) = 0.033; \ \ p_{\text{flip}}(l = 1) = P(C = 0 | R = 1) = 0.3913
\end{equation}
\begin{equation}
    p_{\text{retain}}(l = 0) = P(C = 0 | R = 0) = 0.0.996; \ \ p_{\text{retain}}(l = 1) = P(C = 1 | R = 1) = 0.6087
\end{equation}

\subsection{More Samples for Interpretability}
\label{app-interpret}

\begin{figure}
\centering
\begin{subfigure}{.33\textwidth}
  \centering
    \includegraphics[width=\linewidth]{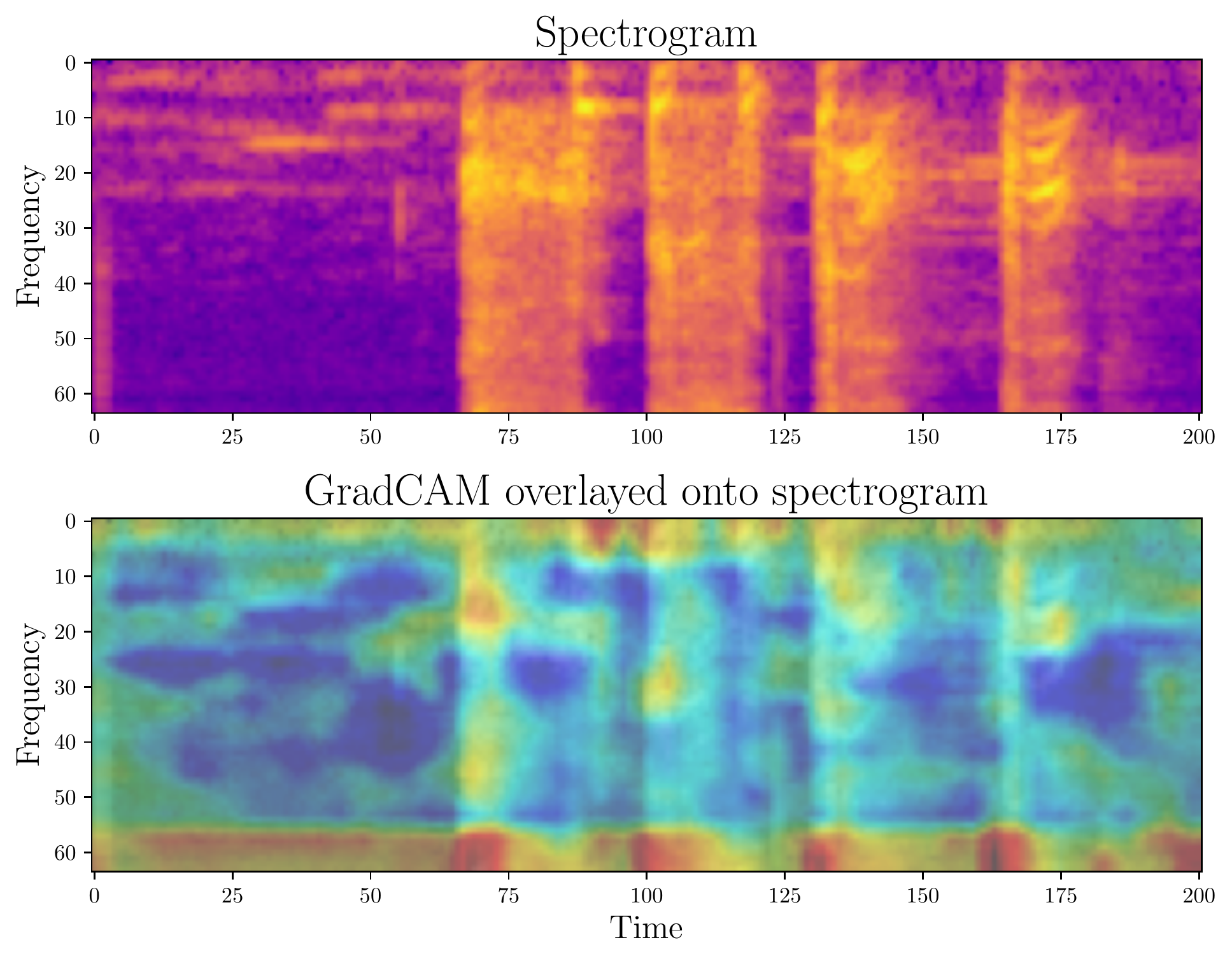}
\end{subfigure}%
\begin{subfigure}{.67\textwidth}
  \centering
    \includegraphics[width=\linewidth]{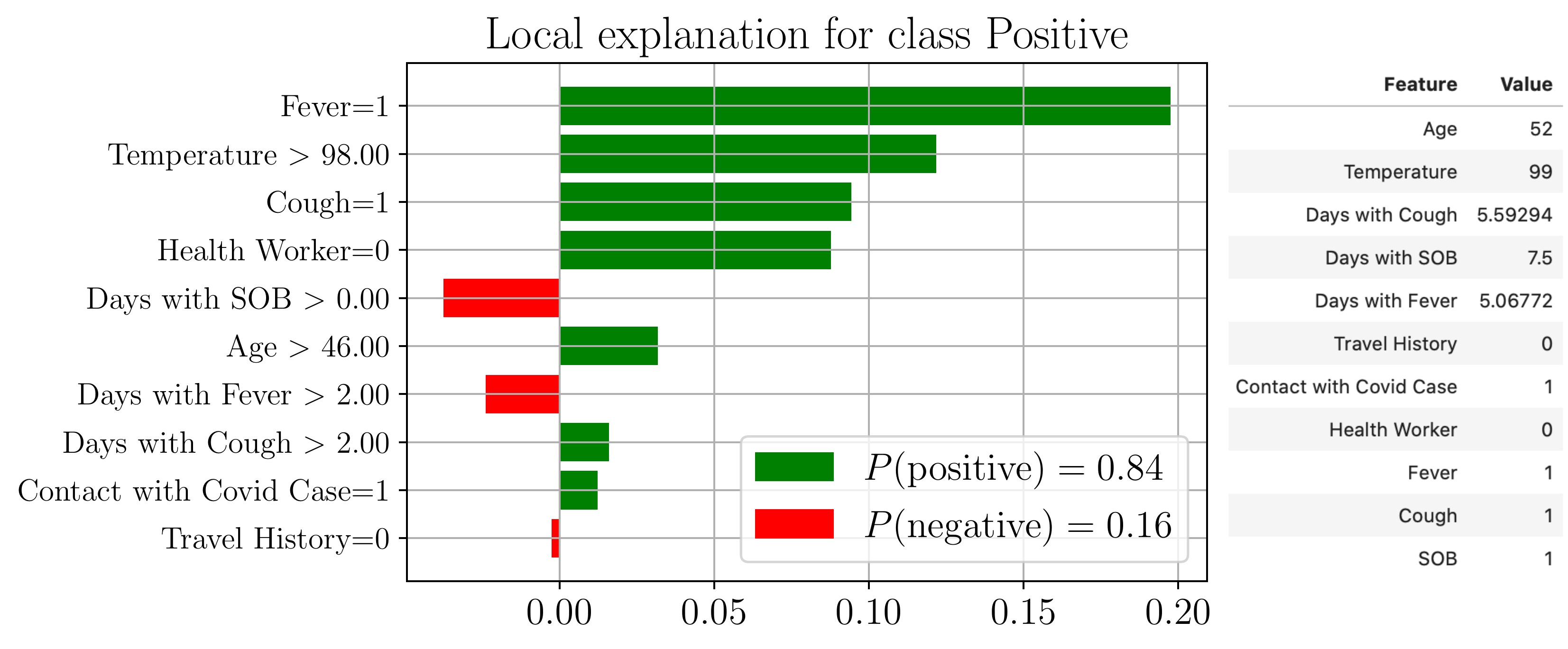}
\end{subfigure}

\centering
\begin{subfigure}{.33\textwidth}
  \centering
    \includegraphics[width=\linewidth]{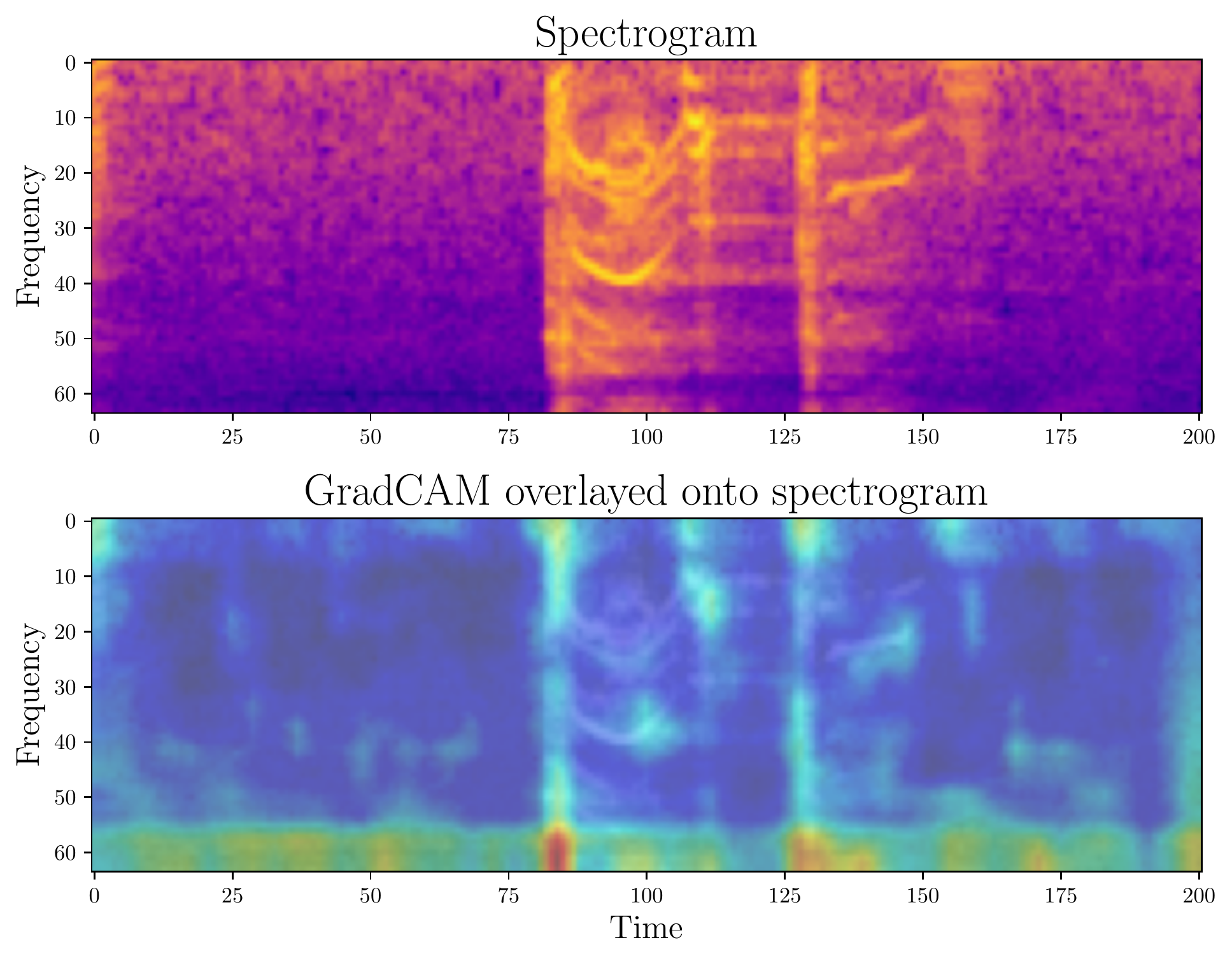}
\end{subfigure}%
\begin{subfigure}{.67\textwidth}
  \centering
    \includegraphics[width=\linewidth]{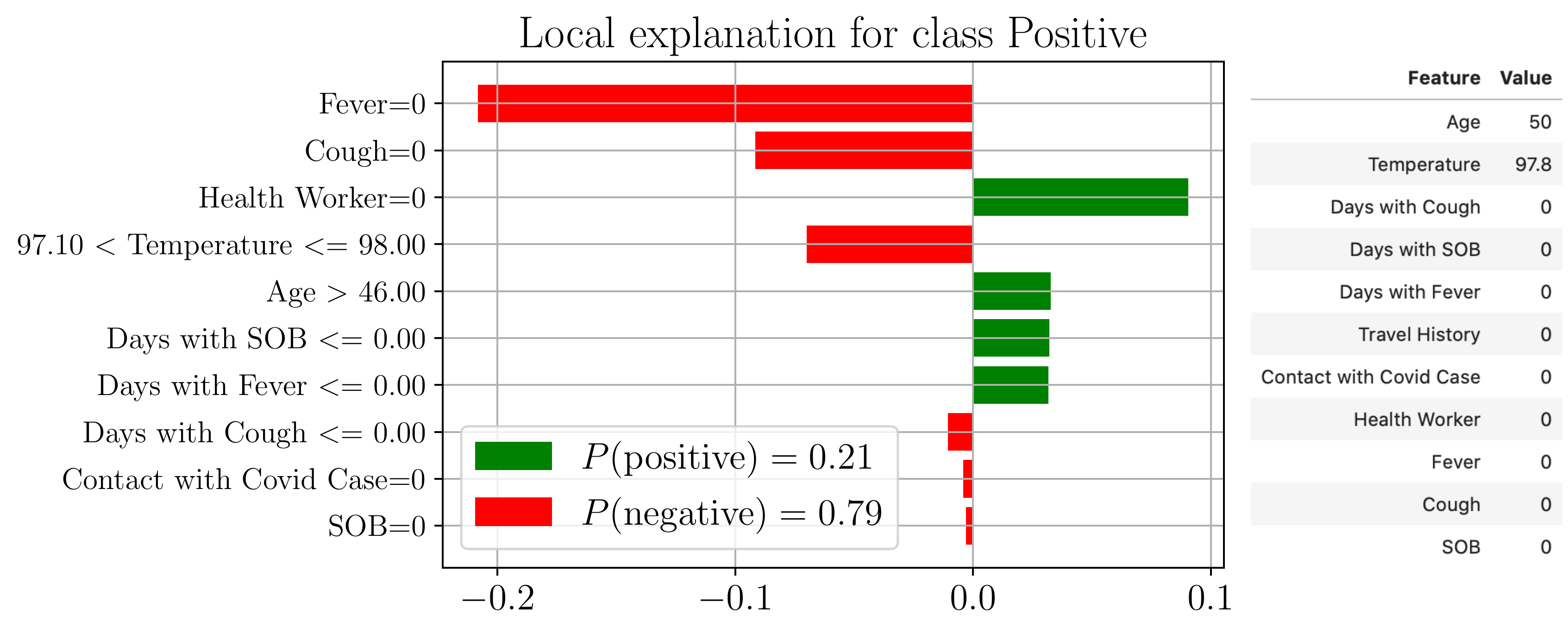}
\end{subfigure}

\centering
\begin{subfigure}{.33\textwidth}
  \centering
    \includegraphics[width=\linewidth]{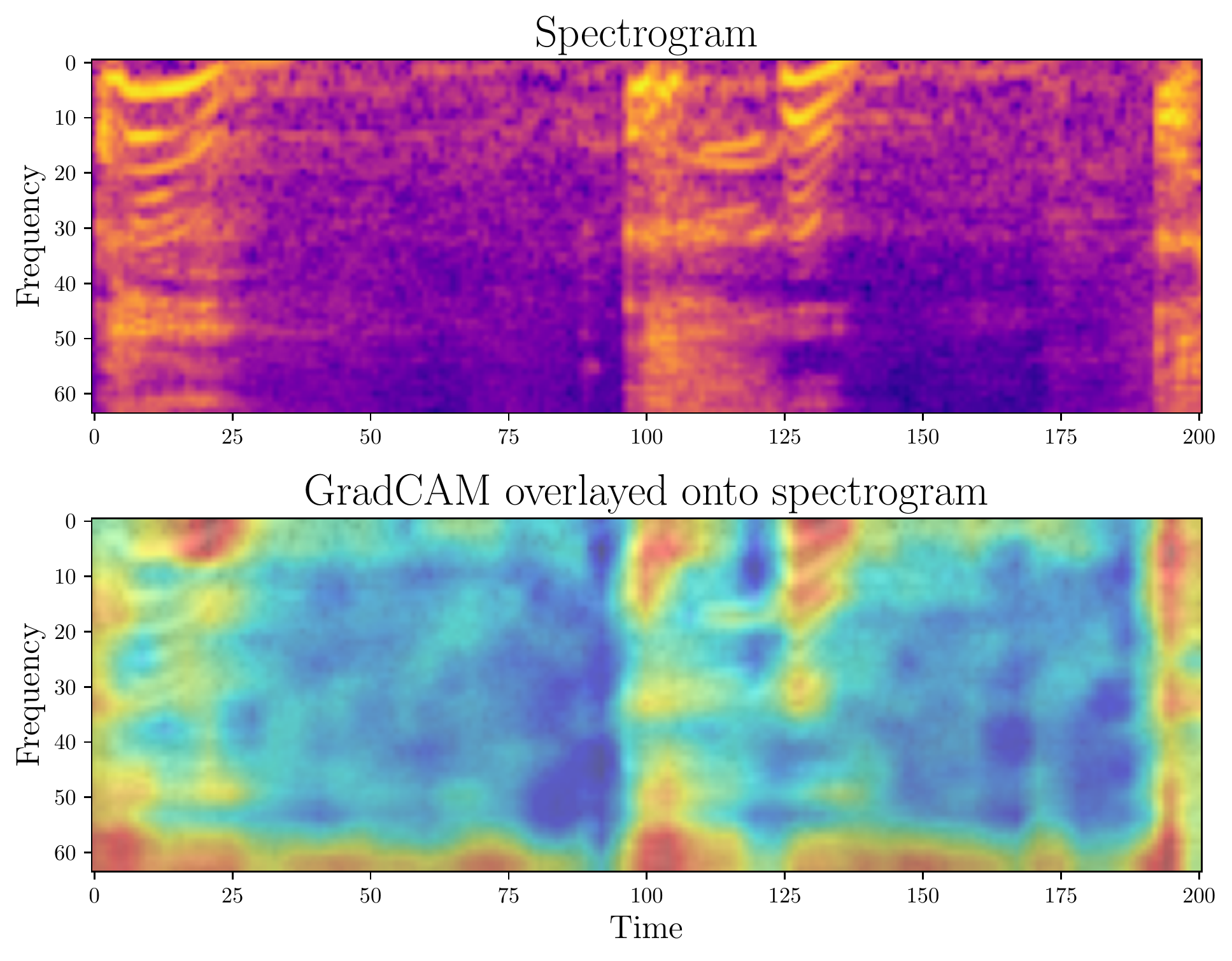}
\end{subfigure}%
\begin{subfigure}{.67\textwidth}
  \centering
    \includegraphics[width=\linewidth]{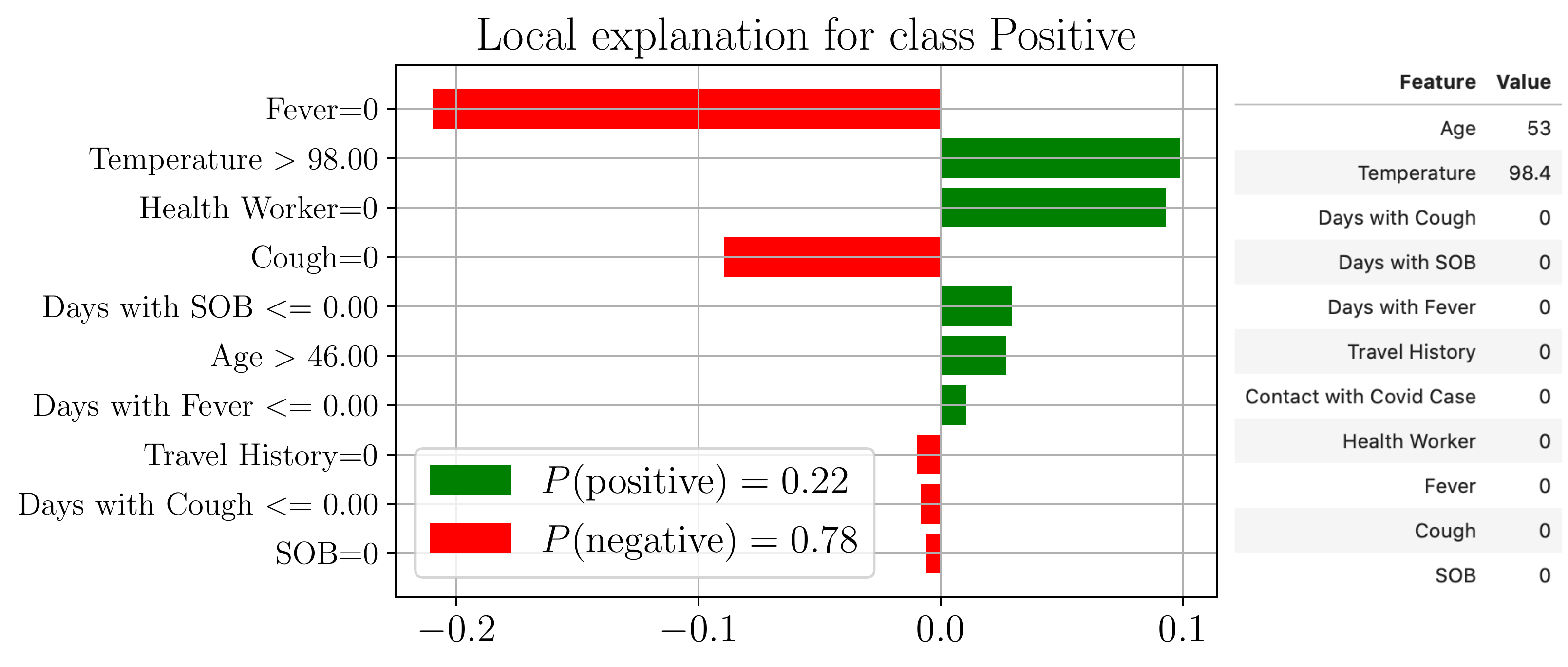}
\end{subfigure}

\centering
\begin{subfigure}{.33\textwidth}
  \centering
    \includegraphics[width=\linewidth]{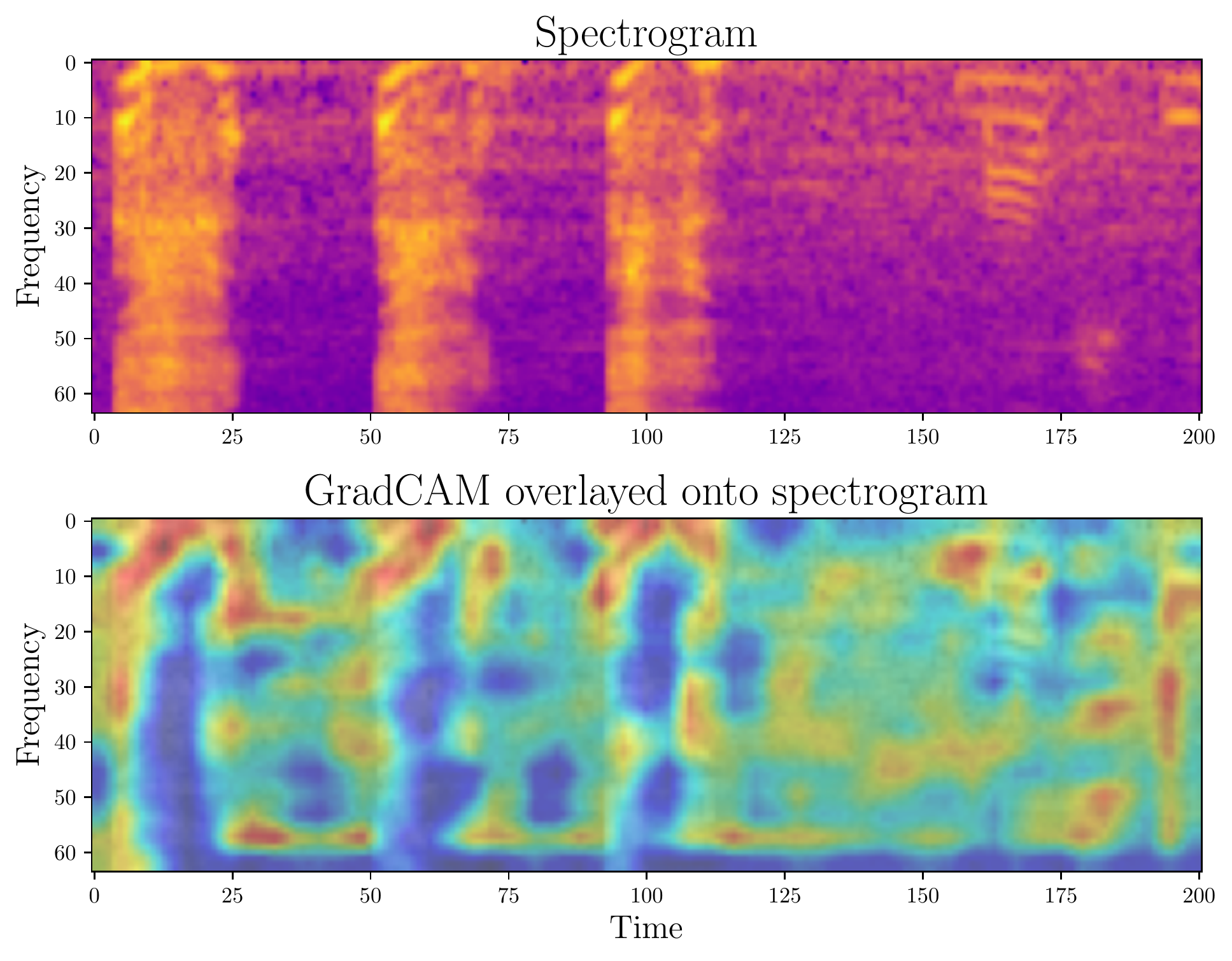}
\end{subfigure}%
\begin{subfigure}{.67\textwidth}
  \centering
    \includegraphics[width=\linewidth]{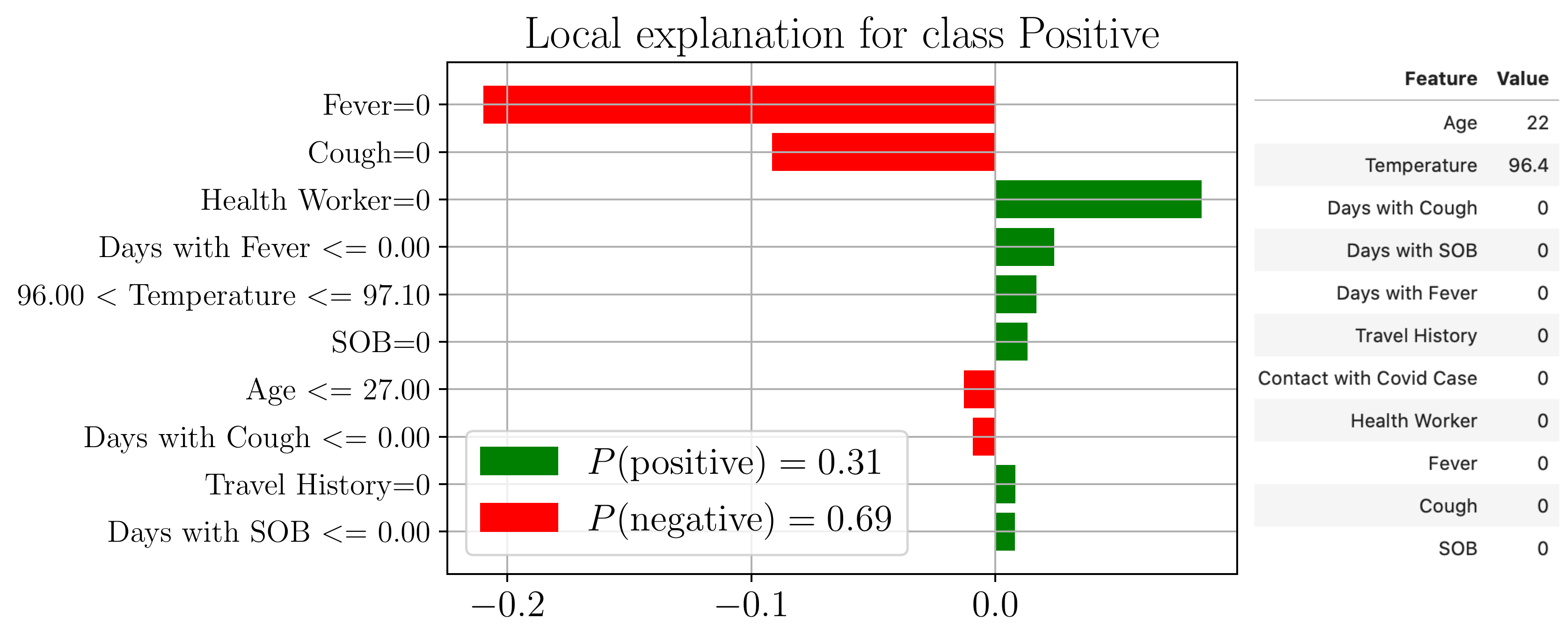}
\end{subfigure}
\caption{More examples of interpretability analysis: (Left) GradCAM++ saliency maps overlaid onto spectrograms. (Right) Contribution of contextual feature to prediction scores based on LIME method. The left and right illustrations are for the same individuals in each row.}
\end{figure}

\end{document}